# MultiLock: Mobile Active Authentication based on Multiple Biometric and Behavioral Patterns

Alejandro Acien, Aythami Morales, Ruben Vera-Rodriguez, and Julian Fierrez, Member IEEE
Universidad Autonoma de Madrid, 28049 Spain
{alejandro.acien, aythami.morales,ruben.vera, julian.fierrez}@uam.es

**Abstract**

*In this paper we evaluate mobile active authentication based on an ensemble of biometrics and behavior-based profiling signals. We consider seven different data channels and their combination. Touch dynamics (touch gestures and keystroking), accelerometer, gyroscope, WiFi, GPS location and app usage are all collected during human-mobile interaction to authenticate the users. We evaluate two approaches: one-time authentication and active authentication. In one-time authentication, we employ the information of all channels available during one session. For active authentication we take advantage of mobile user behavior across multiple sessions by updating a confidence value of the authentication score. Our experiments are conducted on the semi-uncontrolled UMDAA-02 database. This database comprises smartphone sensor signals acquired during natural human-mobile interaction. Our results show that different traits can be complementary and multimodal systems clearly increase the performance with accuracies ranging from 82.2% to 97.1% depending on the authentication scenario.*

## 1. Introduction

Services are migrating from the physical to the digital domain in the information society. Examples can be found in e-government, banking, education, health, commerce, and leisure. This digital revolution is associated with a massive deployment of mobile devices including multiple sensors (e.g. camera, gyroscope, GPS, touch screens, etc.), and full connectivity (e.g. bluetooth, WiFi, 4G, etc.). The mobile market has expanded to the point where the number of mobile devices in use is nearly equal to the world's population. Mobile devices are rapidly becoming data hubs, used to store e-mail, personal photos, online history, passwords, and even payment information. Recent studies have shown that about 34% or more users did not use any form of authentication mechanism on their devices [1]. In similar studies, inconvenience is always shown to be one of the main reasons why users do not use any authentication mechanism. In [2], researchers showed that mobile device users spent up to 9% of the time they use their smartphone unlocking the screens, and the 2018 Meeker Report indicated that the average smartphone user checks his device 150 times per day. Those factors lead individuals to make less security conscious decisions like leaving their smartphones unprotected or just protecting them using simple to break authentication mechanisms (e.g., simple Google unlock graphical patterns vulnerable to over-the-shoulder attacks [3]).

Biometric technologies improve in several ways traditional recognition technologies based on passwords or swipe patterns. The advantages of biometric systems are many in terms of security and convenience of use, which has led these technologies to take on a leading role in the last years. In fact, there is a growing interest in the biometrics research community towards more transparent and robust authentication methods that make use of the interaction signals originated when using the smartphone [4][5]. Signals generated with the sensors already embedded in mobile devices (e.g., gyroscope, magnetometer, accelerometer, GPS, and touchscreen interactions) along with metadata associated to our use of the technology (e.g. internet point access, browsing history, app usage) could assist in user authentication avoiding the inconveniences of traditional unlocking systems. All this information is originated naturally during the normal usage of the user with a smartphone, and it has been demonstrated that can be used for person identification under certain conditions [5]. By regularly conducting unobtrusive identity checks of the mobile user during a normal session, a continuous authentication system can verify if the device is still being operated by the authorized user. With this active system, if the mobile device is stolen, it should quickly recognize the presence of an unauthorized user.

The aim of this paper is to analyze multi-modal approaches to improve the performance of mobile authentication. Our experiments include up to four different biometric traits (touch gestures, keystroking, gyroscope, and accelerometer) and three behavioral-based profiling techniques (GPS, WiFi, and app usage). The experiments are conducted on the UMDAA-02 mobile database [6], a challenging dataset acquired under natural conditions.

Previous works have demonstrated the potential of biometric and behavioral-based profiling patterns for user authentication under controlled scenarios. However, the performance of biometric mobile authentication based on human interaction raises doubt under challenging non-supervised scenarios. The contributions of this work are: i) performance analysis of user authentication based on 4 biometric data channels (touch gestures, keystroking, accelerometer, and gyroscope) and 3 behavior profiling data sources (WiFi, GPS, and App usage), obtained during natural human-smartphone interaction; ii) study of multimodal approaches for smartphone user authentication based on various combinations of the previous 7 data channels, both for One-Time Authentication and for Active Authentication schemes (i.e., continuously over multiple sessions).

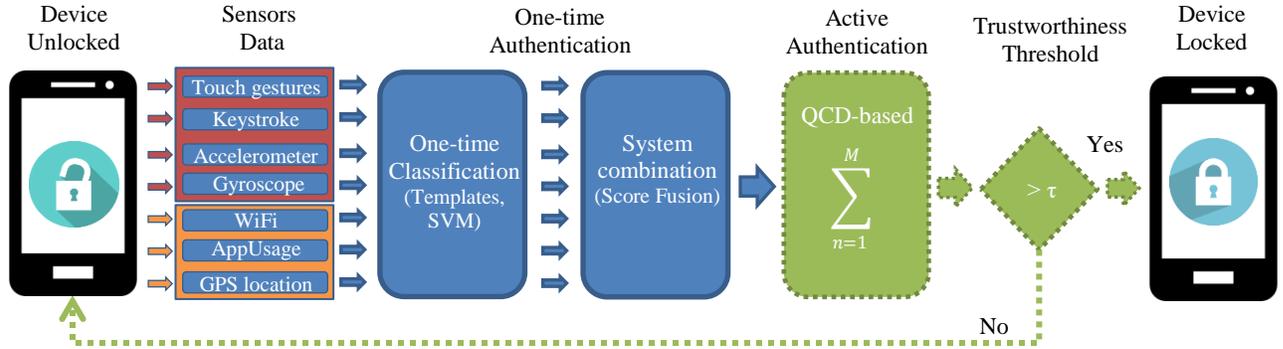

**Figure 1:** System architecture. Continuous line corresponds to one-time authentication, and dotted line indicates add-on modules for active authentication.

The rest of this paper is organized as follows: Section 2 links the present works with related research. Section 3 describes the architecture of our approach. Section 4 explains the experimental protocol, describing the database and the experiments performed. Section 5 presents the final results for single and multimodal architecture and Section 6 summarizes the conclusions and future work.

## 2. Related Work

Mobile authentication based on soft biometrics traits has been deeply studied in the last years [7][8][9]. Swipe dynamics is one of the most popular traits analyzed [7]; however, it has been shown not to have enough discriminative power to replace traditional technologies.

Accelerometer and gyroscope sensors have been studied traditionally for gait recognition, and some works have demonstrated also their utility for user authentication with acceptable performance [10].

Geo-location based verification approaches are scarce in the literature. In [11], Mahbub and Chellappa developed a mobile authentication system using trace histories by generating a confidence score of the new user location taking into account the sparseness of the geo-location data and past locations. For this purpose, they employed modified Hidden Markov Models (HMMs) considering the human mobility as a Markovian motion. In a similar way, in [12] a variation of HMMs was used to develop a user authentication mobile system by exploiting application usage data. They suggest that unforeseen events and unknown applications have more impact in the authentication performance than the most common apps used by the user.

The potential of WiFi history data was analyzed in [8] for mobile authentication. They explored: i) the WiFi networks detected by the smartphone, ii) when the detection occurs, and iii) how frequently those networks are detected during a period of time.

Regarding keystroke traits, in [9] a fixed-text keystroking system for mobile user authentication was studied using not only time and space based features (e.g. hold and flight times, jump angle or drag distance) but also studying the hands postures during typing as discriminative information. In [13], a novel fixed-text authentication system for laptops and mobile devices based on Partially Observable HMMs was studied. This model is an extension of HMMs in which the hidden state is conditioned on an independent Markov chain. The algorithm is motivated by the idea that typing events depend both on past events and also on a separate process.

Finally, building a multimodal system that integrates all these heterogeneous information sources for mobile user authentication is still a challenge [14]. Noisy data, intra class variation or spoofing attacks [15] are some inevitable problems in unimodal systems that can be overcome by multimodal architectures [5][14]. In [16], a multimodal user authentication system was based on the fusion at decision level of voice, location, multi-touch, and accelerometer data. Their preliminary results suggest that these four modalities are suitable for continuous authentication. In [17], a fusion was performed also at decision level of behavioral-based profiling signals such as web browsing, application usage, and GPS location with keystroking data achieving 95% of user authentication accuracy using information from one-minute window. More recently, in [18] a fusion also at decision level of touch dynamics, power consumption, and physical movements modalities achieved 94.5% of accuracy with a dataset that was captured under supervised conditions. In [19], an unobtrusive mobile authentication application is designed for single and multimodal approaches. They collected data from WiFi, Bluetooth, accelerometer, and gyroscope sources in unsupervised conditions and fused them at score level achieving up to 90% of accuracy in the best scenario.

**TABLE 1:** Example of an app-usage user template generated according the data captured during six days.

| Event | Time slot | Frequency |
|---|---|---|
| WhatsApp | 4 | 5 |
| Navigator | 4 | 3 |
| YouTube | 5 | 1 |
| WhatsApp | 5 | 1 |
| Facebook | 7 | 2 |

## 3. System Description

In this paper we analyze 4 biometric data channels (touch gestures, keystroking, gyroscope, and accelerometer) and 3 behavior data sources (GPS, WiFi, and app usage). We study 2 architectures for user authentication (see Figure 1): the first approach (continuous line in Figure 1), referred to as One-Time Authentication (OTA) is based on unimodal systems trained with the information extracted from the mobile sensors during a user session. A session is defined as the elapsed period between the device unlock and the next lock. Therefore, sessions have a variable duration and information obtained from sensors varies depending on the usage of the device during the session. The information provided by the sensors is employed to model the user according to the seven systems mentioned before: keystroking, touch gestures, accelerometer, gyroscope, WiFi, app usage, and GPS location. Each system provides a single authentication score and these scores are combined to generate a unique score for each session. The second approach, called Active Authentication (dotted line in Figure 1), is based on updating a confidence value generated from the One-Time Authentication during consecutive sessions.

The seven systems are categorized into two main groups according to the nature of the information employed to model the user: biometric and behavior-based profiling systems. In this work, biometric systems refer to the top 4 channels in the Sensors Data module of Figure 1 (red box). The way we realize touch gestures, typing, or handle the device is determined by behavioral aspects (e.g. emotional state, attention) and neuromotor characteristics of users (e.g. ergonomic, muscles activation/deactivation timing, motor abilities). Behavioral-based profiling refers to those systems that model the owners of the device according to the services they use during their daily habits (orange box in Figure 1, bottom 3 channels in the Sensors Data module).

### 3.1. Behavioral-based profiling systems

WiFi, app usage, and GPS location system are based on a similar template-based matching algorithm. A user template is defined as a table containing the time stamps and the frequency of the events [8]. For this, we divided the time (24 hours of the day) into $N$ equal time slots (e.g. if we choose $N = 48$ we will have 48 time slots of 30 minutes), giving to each time slot a number ID. Then we store in the template the event`s name, the number ID of the time slot and the occurrence frequency of that event (number of times this event occurs during this particular time slot on a window of consecutive days). Table 1 shows an example of the app-usage template for a given user generated according the data obtained during six days; in this case *WhatsApp* application is detected in the fourth slot for five days out of the six days considered meanwhile the same app is detected only one day in the fifth slot. Note that multiple detections of the same event in the same time slot and day are ignored but they are stored if they belong to different time slots or days. Depending on the system, the event could be the name of the WiFi network, latitude and longitude of a location (with two decimals of accuracy), or the name of a mobile app for WiFi, GPS location, and app usage systems, respectively.

Finally, we test the systems by calculating a behavior-based confidence score [8] for each test session as:

$$score = \sum_{i=1}^{S} f_i^2 \quad (1)$$

where $f_i$ is the frequency of the event stored in the template that match with the test event $i$ in the same time slot and $S$ is the total number of events detected in that test session. For example, if the test session includes the usage of *WhatsApp* and *Navigator* apps during the fourth slot, the score confidence will be $5^2 + 3^2 = 34$ (according to the template showed in Table 1). Based on this, a higher score in the test session implies higher confidence for authentication.

### 3.2. Biometric systems

For touch gestures, keystroking, accelerometer and gyroscope systems, the feature extraction and classification algorithms are adapted to model the user information. In the touch gestures system, the feature set employed is a reduced set of the global features presented in [20] (commonly used for online handwriting sequence modeling) and adapted for swipe biometrics in [7]. Mean velocity, max acceleration, distance between adjacent points, or total duration are some examples of this subset of 28 features extracted (see [20] for details).

For accelerometer and gyroscope, the data captured comprises *x*, *y*, and *z* coordinates of the inclination vector of the device (gyroscope) and the acceleration vector (accelerometer) in each time stamp. For these 2 sensors we use the feature set proposed in [10]: mean, median, maximum, minimum, distance between maximum and minimum, and the standard deviation for each array of

**TABLE 2:** General UMDAA-02 dataset information.

| Description | Statistics |
|---|---|
| Gender | 36M/12F |
| Age | 22-31 years |
| Avg. Days/User | 10 days |
| Avg. Sessions/User | 248 sessions |
| Avg. Time/Session | 224 seconds |
| Avg. Systems/Session | 5.2 systems* |

*Systems: refers to the number of systems available out of the 7 studied in this work.

coordinates. Moreover, we propose the 1 and 99 percentiles and the distance between them as additional features.

Regarding keystroking, the keys pressed were encrypted in order to ensure user-privacy so systems based on graphs were discarded and we adopted traditional timing features: hold time, press-press latency, and press-release latency as in [21][22]. Finally, we propose a feature set based on six statics (mean, median, standard deviation, 1 percentile, 99 percentile, and 99-1 percentile). Note that UMDAA-02 keystroke data can be considered as a free text scenario. However, the limited samples per session and the encrypted keys difficult the application of popular free-text keystroke authentication methods.

For classification we train different Support Vector Machines (SVM) with a radial basis function (RBF) kernel, one for each feature set and user with an optimization of both hyperparameters ($C$, $\sigma$).

## 4. Experiments

### 4.1. Database

The experiments were conducted with UMDAA-02 database [6]. This database comprises 141.14 GB of smartphone sensor signals collected from 48 Maryland University students over a period of 2 months, the participants used a smartphone provided by the researchers as their primary device during their daily life (unsupervised scenario). The sensors captured are touchscreen (i.e. touch gestures and keystroking), gyroscope, accelerometer, magnetometer, light sensor, GPS, and WiFi, among others. Information related to mobile user´s behavior like lock and unlock time events, start and end time stamps of calls and app usage are also stored.

Table 2 summarizes the characteristics of the database. During a session, the data collection application stored the information provided by the sensors in use.

### 4.2. Experimental Protocol

The experiments are divided into two different scenarios: One-Time Authentication (OTA) and Active Authentication (AA). In OTA the performance is calculated using only one session to authenticate the user meanwhile in AA we employ multiple consecutive sessions in order to improve the confidence in the authentication.

For all experiments the dataset is divided into 60% days for training (first sessions) and the remaining 40% days for testing. This means that we employ six days in average to model the user and 4 days in average to test such a model. The performance for both scenarios is presented in terms of average correct classification rate computed as 100−EER (Equal Error Rate). EER refers to the value where False Acceptance Rate (percentage of impostors classified as genuine) and False Rejection Rate (percentage of genuine users classified as impostors) are equal.

**4.2.1. One-Time Authentication.** In OTA experiments, all 7 systems are trained separately for each user and the scores are calculated at session level, generating 7 scores for each test session as maximum (note that the number of systems available during a session varies). The 4 biometric systems considered can produce more than one score per session (e.g. multiple gestures or multiple keystroking sequences during a text chat). In those cases, the scores available during the session are averaged to obtain one score for each biometric system and session. Finally, we normalize with $tanh$ normalization and fuse the scores (mean rule) to calculate a single score [14] according to the different fusion set-ups proposed. The scores from the best fusion set-up will be used in the AA scenario.

**4.2.2. Active Authentication.** For AA experiments we consider the QCD algorithm (Quickest Change Detection) as explained in [23]. The QCD-based algorithm updates a confidence score based on previous events (sessions in this work) by performing a cumulative sum of scores. This cumulative sum will be almost zero if the scores belong to the genuine user, and will grow if an impostor takes the control, until it reaches a certain threshold that would detect the intruder. The cumulative sum is calculated as follow:

$$score_j^{AA} = \max(score_{j-1}^{AA} + L_j, 0) \quad (2)$$

where $j$ means the actual session and $score_{j-1}^{AA}$ is the previous cumulative score. $L_j$ is the contribution of the actual session calculated as the log-likelihood ratio between score distributions:

$$L_j = \log\left(\frac{f_I(score_j)}{f_G(score_j)}\right) \quad (3)$$

where $f_G$ and $f_I$ are the probability distributions of the genuine and impostor scores respectively calculated previously in the OTA fusion scenario, and $score_j$ is the OTA fused score of the actual session. According to (3), the log-likelihood ratio $L_j$ will be negative if $score_j$ belongs to a genuine user and positive in the opposite case and, therefore, multiple consecutive sessions of an impostor in control will increase the cumulative sum ($score_j^{AA}$).

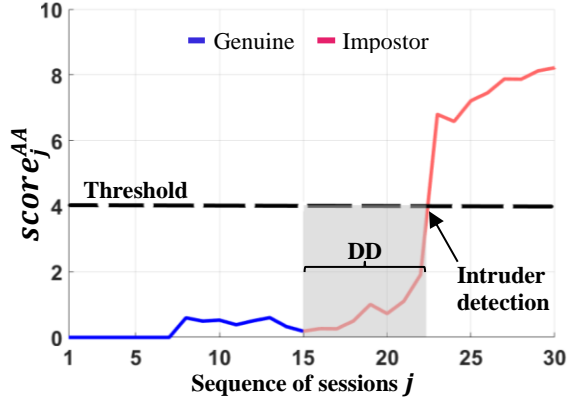

**Figure 2:** An example of QCD-based curve with a sequence of 30 sessions (15 genuine and 15 impostors). The dashed line is the intruder detection threshold and the grey box shows the Detection Delay (DD).

Figure 2 depicts an example of $score_j^{AA}$ evolution. At the time the mobile starts to be operated by an intruder (session number sixteen in Figure 2) the $score_j^{AA}$ ($j > 16$) will tend to increase until reaching the threshold. The time elapsed between the intrusion start and the intrusion detection is known as Detection Delay (DD) measured in number of sessions.

## 5. Results and Discussion

### 5.1. One-Time Authentication

In this section we analyze the OTA scenario: the accuracy for the 4 biometric systems and the fusion with each behavior-based profiling system. Table 3 summarizes the final results by ranking from the best individual biometric system performance to the worst one. The first column shows the performance obtained for each single biometric system. From the second to the fourth column, we show the performance for the fusion of each biometric system with each behavior-based profiling system, and the fifth column shows the fusion with all of them.

First of all, the poor performance achieved by some biometric systems can be caused by the uncontrolled acquisition conditions of the database and the limited number of samples per session (e.g. free text keystroke usually requires large sequences) but the combination of all of them (last row in Table 3) shows acceptable performance for unsupervised scenarios.

Secondly, we can observe that behavior-based profiling systems always improve biometric systems performances in all fusion schemes. In fact, the combination of all behavior-based profiling approaches with each biometric system achieves the most competitive performance, improving them in more than 18% of accuracy in the best of cases. If we analyze each single behavior-based profiling fusion, we can observe that the GPS system achieves the best improvements, boosting biometric systems performances in more than 13% of accuracy.

Finally, in Figure 3 we plot the ROC curves for each single biometric system and the best fusion set-up: the fusion of all behavior-based profiling systems with each biometric system (column 5 in Table 3). The results in OTA scenario suggest that behavior-based profiling systems always improve the biometric ones and the best performance is achieved by fusing with all of them, and therefore, the scores obtained from this fusion scheme will be use in AA scenario.

### 5.2. Active Authentication

First some definitions:

*Probability of False Detections (PFD):* is the percentage of genuine users detected as intruder during a sequence of genuine sessions. It means that $score_j^{AA}$ reaches the intruder detection threshold during a genuine session sequence (genuine curve in Figure 2). PFD is similar to FMR (False Match Rate) in one-time authentication.

*Probability of Non-Detection (PND):* is the percentage of intruders not detected during a sequence of intruder sessions. It means that $score_j^{AA}$ does not reach the intruder detection threshold during the intruder sessions sequence (impostor curve in Figure 2). PND is similar to FNMR (False Non-Match rate) in one-time authentication.

*Average Detection Delay (ADD):* is the average number of impostor sessions needed to detect an intruder (the grey box in Figure 2).

**TABLE 3:** Results achieved for both One-Time and Active Authentication (AA) scenarios in terms of correct classification rate (%) according to different number of biometric systems and their fusion with behavior-based profiling systems. In brackets, average number of sessions employed (ADD).

| System | Acc. | +WiFi | +GPS | +AppUsage | All | AA |
|---|---|---|---|---|---|---|
| Touch gestures | 72.0 | 78.2 | 78.3 | 75.4 | **83.1** | 95.0 (6) |
| Keystroking | 62.5 | 72.6 | 70.9 | 67.8 | **79.1** | 92.9 (7) |
| Accelerometer | 61.3 | 70.8 | 77.3 | 64.7 | **78.7** | 93.7 (7) |
| Gyroscope | 59.5 | 69.7 | 72.6 | 63.4 | **78.4** | 92.3 (6) |
| Combined | 73.2 | 77.3 | 78.9 | 75.3 | **82.2** | **97.1 (5)** |

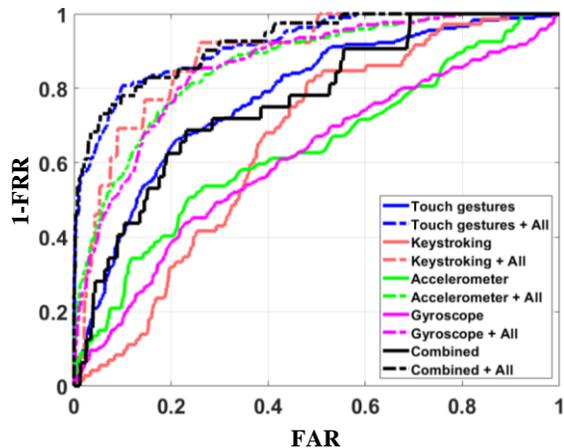

**Figure 3:** ROC curves (One-Time Authentication) for individual biometrics and the best fusion set-up incorporating the 3 considered behaviour profiling sources (All = WiFi + GPS + App usage).

To calculate the correct classification rate in AA we plot in Figure 4 the PND vs. PFD and ADD vs. PFD curves. The PND-PFD curves are similar to FMR-FNMR curve in one-time authentication with the main difference that those results are obtained from a sequence of stacked scores instead of only one. The equal error rate (EER) will be the value where PND and PFD are equal and the correct classification rate will be computed as $100 - EER$. The ADD-PFD curve shows the number of sessions needed to detect an intruder according to the PFD. This curve allows us to know how many sessions are needed to achieve the EER reported. For instance, the PND-PFD curves in Figure 4 (right) show that the EER in Active Authentication is 2.9% and the ADD to achieve that EER is 5 sessions. This means that we can improve OTA results at the cost of having more sessions to detect an intruder. All curves were calculated for each user and averaged.

Finally, all AA results are summarized in the last column of Table 3. Remember that scores employed in the QCD-based algorithm come from the fusion scores of the best OTA scenario (fusing with all behavior-based profiling systems) so both performances are correlated. Each performance in Table 3 for AA is followed by the average detection delay in brackets needed to achieve it. As we expected, in all different fusion set-ups the AA algorithm improves the accuracy at the cost of needing more sessions to detect the intruder. In fact, for the best fusion set-up the performance improves from 82.2% to 97.1% by using 5 consecutive intruder sessions to detect the impostor. Comparing all scenarios, the greatest improvement occurs with all biometric systems combined (14.9% of improvement in the last row of Table 3) with an average 5 sessions.

## 6. Conclusions and Future Work

In this paper, we have studied user mobile active authentication based on multiple biometric and behavior-based profiling systems. For this, we studied two scenarios according to the number of sessions used: one session (One-Time Authentication) and multiple sessions (Active Authentication). The results suggest that some biometric systems work better than others, and that the fusion with behavior-based profiling systems always improves the results, achieving accuracies up to 82.2% in the best case for an OTA scenario. Our experiments also suggest that Active Authentication always improve the accuracies with up to 14% of enhancement with respect to One-Time Authentication using between 5 and 7 sessions.

For future works we will work to improve the performance of individual systems, especially biometrics systems. Better individual performances will produce better fused schemes. The combination of heterogeneous data at data and feature level will be evaluated in order to merge correlations between systems (e.g. touch gestures and apps used are highly correlated).

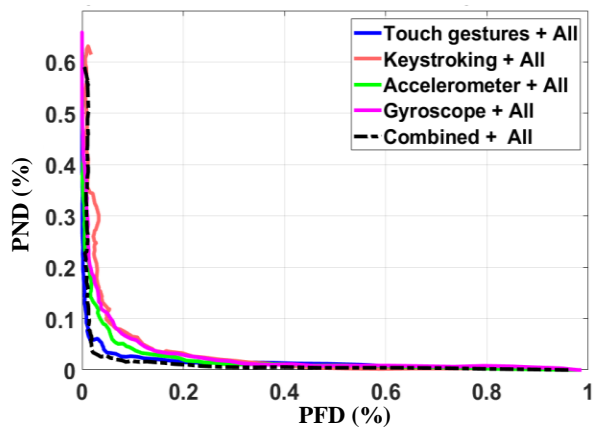
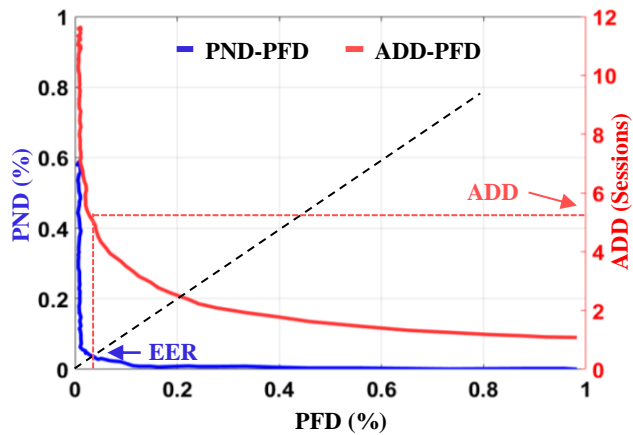

**Figure 4:** PND vs PFD curves of active authentication for the best fusion schemes (left), PND vs PFD and ADD vs PFD curves for the best fusion set-up (right). The dark dashed line shows the EER and the red one shows the Average Detection Delay for that EER in the right plot.